\documentclass[preprintnumbers,amsmath,floatfix,11pt]{revtex4}
\usepackage{bm}

\begin{document}
\preprint{FERMILAB-CONF-07-50-T}
\vspace*{0.7in}

\title{Weak Mixing and Rare Decays in the Littlest Higgs Model}

\author{William A. Bardeen}

\affiliation{
Theoretical Physics Department\\
Fermilab, MS 106, P.O. Box 500\\
Batavia, IL 60510}
\date{March 15, 2007}

\begin{abstract}
Little Higgs models have been introduced to resolve the fine-tuning
problems associated with the stability of the electroweak scale and
the constraints imposed by the precision electroweak analysis of
experiments testing the Standard Model of particle physics.  Flavor
physics provides a sensitive probe of the new physics contained in
these models at next-to-leading order. 
\end{abstract}

\maketitle

\section{Introduction}
The Standard Model has been extraordinarily successful in
describing all known phenomena of particle physics with the
possible exception of the astrophysical evidence for dark matter.
Precision experiments including the results from LEP, the Tevatron
and the B-factories have confirmed detailed predictions of the
Standard Model and have placed strong constraints on new physics
associated with the electroweak scale.  Flavor changing processes
such as weak mixing and neutral current processes in rare decays
are strongly suppressed in the Standard Model and provide a unique
window on new physics at scales much above the electroweak scale. 

Despite the success of the Standard Model, there remain puzzles
concerning the mechanisms that determine parameters of the SM which
must be finely adjusted to agree with experiment.   These include the
theta angle of strong CP violation, the Higgs mass term which
determines the electroweak scale and the net vacuum energy density
which may be responsible for the dark energy inferred from
astrophysical data. 
In particular, the Higgs mass term gets large radiative corrections
from Standard Model processes which would normally be expected to
destabilize the electroweak scale.  If new physics is introduced to
stabilize these radiative corrections, then there can be a tension
between the SM fits to precision electroweak data and the new physics
contributions. 

\section{The Littlest Higgs Model}
Little Higgs models \cite{1,2} are based on the dynamics of
pseudo-Nambu-Goldstone bosons where the Higgs mass term is protected
from radiative corrections by the existence of new symmetries that are
dynamically broken.   In the absence of additional explicit breaking
of these new symmetries, the physical Higgs boson remains 
massless \cite{3}.  The Higgs mass will be protected from the quadratically
divergent radiative corrections if the explicit symmetry breaking
occurs collectively, ie. two or more symmetry breaking terms must be
present in order to generate a mass for the physical Higgs boson.   

The main ingredients of Little Higgs models are an extended
electroweak gauge group, $G_1\otimes G_2 \to$SM, which is embedded in
a larger global 
symmetry with dynamical symmetry breaking, and an extended top quark
sector needed to accommodate the observed top quark mass.  The global
symmetries are broken collectively and the physical Higgs boson is a
pseudo-Nambu-Goldstone boson.  There are typically three important
scales in Little Higgs models,  $\Lambda$ ($10\to 30$ TeV) -- the
scale of new dynamics and the 
effective cutoff,  $f$ ($1\to 3$ TeV), $f\sim \Lambda/4\pi$ -- the
scale of dynamical breaking of the global 
symmetries and the mass scale of the new particles used to cancel the
quadratic divergences, and  $v$ (175 GeV), $v\sim f/4\pi$ -- the
electroweak scale and the mass of 
the SM Higgs boson, the SM gauge bosons and the top quark. 

The Littlest Higgs model \cite{4} is based on the Nambu-Goldstone bosons of
the coset space of $SU(5)/SO(5)$.  The gauge boson dynamics is based on the
$[SU(2)\otimes U(1)]_1 \otimes [SU(2)\otimes U(1)]_2$ subgroup of
$SU(5)$ which is broken to the electroweak gauge group, $(SU(2)\otimes
U(1))_{\rm SM}$, at the scale $f$.  Four of the fourteen
Nambu-Goldstone bosons are absorbed by 
the heavy gauge bosons, $(W_H^{\pm}, Z_H^0, A_H)$.  Of the ten
remaining Nambu-Goldstone 
bosons, six form a complex triplet $(\Phi^{++}, \Phi^+, \Phi^0)$ and
acquire a mass of order $f$, and 
four form a complex doublet $(H^+, H^0)$ corresponding to the usual
Higgs field of 
the Standard Model.  The top quark sector is also modified.  A heavy
vector-like top quark is needed in addition to the Standard model top
quark in order to cancel the quadratic divergences associated with top
quark loops.   

In the Littlest Higgs model described above, the heavy gauge bosons
mix with the light SM gauge bosons at tree level. These mixings can
result in significant modifications to the predictions of the Standard
Model that are highly constrained by precision electroweak
measurements.  These constraints will force the symmetry breaking
scale, $f$, to be large \cite{5}.  However, a large scale for $f$ can reintroduce
issues associated with the fine-tuning of the Higgs mass parameters.   

This tension can be relaxed through the introduction of a new
symmetry, T-parity \cite{6}, which prohibits the mixing between the heavy
and light gauge bosons.  The heavy gauge bosons and the complex
triplet scalars are odd under T-parity while the complex doublet Higgs
field and the Standard Model gauge bosons are even.  The top quark
sector must also be modified to preserve T-parity.  Since all
interactions are to be invariant under T-parity, the lightest odd
parity particle will be a dark matter candidate.  The precision
electroweak constraints for the Little Higgs model with T-parity have
been carefully analyzed \cite{6} and lower scales for the new physics can
be tolerated with less tension with the fine-tuning of the Higgs mass
parameters. 

\section{FCNC Processes in the Littlest Higgs Model}
Flavor-changing processes are highly suppressed in the Standard Model
and, therefore, provide a unique opportunity for the exploration of
new physics contributions.  Flavor physics has long been focus of
Andrzej Buras and his collaborators.  I will report on some of their
conclusions concerning weak mixing and flavor-changing neutral current
processes in the Littlest Higgs model with and without T-parity.  The
reader is referred to the original papers for detailed predictions and
specific results. 

Weak mixing amplitudes for $\Delta S=2$ and $\Delta B=2$ processes are described by effective
weak Hamiltonians involving appropriate four-fermion operators with
coefficient functions that are sensitive to the short distance
physics.  At leading order, the coefficient functions are determined
by box diagrams involving both Standard model particles and the new
heavy states of the Littlest Higgs model.  The calculations can be
made in unitary gauge where only physical particles are taken into
account.  Single box diagrams in unitary gauge are divergent but the
divergences all cancel at one loop due to the GIM mechanism.   

For the Littlest Higgs model without T-parity, Buras et al.\cite{7,8} find
that the new physics contributions to $\Delta M_s$, $\Delta M_d$ and $\varepsilon_K$ are all positive.  This implies
a suppression of $|V_{td}|$ and of the angle $\gamma$ in the unitarity triangle as well
as an enhancement of $\Delta M_s$ relative to the Standard Model expectations.  If
the scale $f$ is as small as 1 TeV, the effects amount, at most, to
15-20\% corrections and decrease below 5\% for $f >3$--$4$ TeV as required by the
precision electroweak constraints.  Contributions of the charged
scalars, $\Phi^{\pm}$, turn out to be negligible.  

Rare decays are also described by an effective weak Hamiltonian which
includes both box diagrams and flavor-changing neutral current
processes.  In the Littlest Higgs model without T-parity there is
considerable mixing between the new heavy states and those of the
Standard Model.  In unitary gauge, the analytic results can be
decomposed into six classes of diagrams.  Custodial symmetries are
broken at $O(v^2/f^2)$ in contributions at next-to-leading order. 

The corrections to the coefficient functions for rare decay processes
are at most 15\% for a scale $f\sim O(2\mbox{--}3\mbox{ TeV})$.  The
amplitudes for $K^+\to \pi^+ 
\nu\bar{\nu}$, $K_L\to\pi^0\nu \bar{\nu}$, $B_{s,d}\to\mu^+ \mu^-$ and
$B\to X_{s,d}\nu\bar{\nu}$ are all
calculated\cite{8} but are hard to distinguish from the
Standard Model.  There are also very small corrections
($\sim 4$\%) to the branching ratio for $B\to X_s\gamma$.  The new physics
contributions to rare decays are generally suppressed
by the large scale required for $f$ due to the lack of
custodial symmetry in the Littlest Higgs model without
T-parity. 

A novel feature of this calculation is the presence of residual
ultraviolet divergences that are found in the amplitudes for rare
decay processes at order $(v^2/f^2)$.  These residual divergences are not an
artifact of unitary gauge but are also present in calculations using
renormalizable gauges such as Feynman gauge.  The divergences
represent a true sensitivity to the ultraviolet completion of the
Littlest Higgs model.   

These divergences are very analogous to the renormalization of $G_A$ in a
chiral quark model where chiral loops generate a log divergent
contribution to $G_A$ in the nonlinear version of the theory.  The linear
quark-sigma model is renormalizable and the renormalization of $G_A$ is
finite.  In this case, the cutoff scale in the log divergent term is
replaced by the mass of the scalar partner to the pion and no other
divergences survive. 

In the calculations of the FCNC processes described above, Buras et
al.\cite{8} estimate the divergent terms by using the scale, $\Lambda=4\pi
f$, the
ultraviolet cutoff scale of the Littlest Higgs model.  Finite
corrections could be added reflecting additional ``Leutwyler terms''
in the effective field theory description of the dynamics. 

\section{Flavor Physics in the Littlest Higgs Model with T-parity}
To avoid the strong precision electroweak constraints, Cheng et
al.\cite{6} introduce a T-parity symmetry to the Littlest Higgs model.
The presence of the custodial symmetry in this model allows for a
lower scale $f$ for the new physics and a more realistic solution to the
tension between the Higgs fine-tuning problem and the precision
electroweak constraints.  Hubisz et al.\cite{9} were the first to study
flavor physics and the potential for new physics in weak mixing
processes in the Littlest Higgs model with T-parity.   These processes
were also studied by Buras et al.\cite{10} and extended to FCNC processes
and rare decays \cite{11}.  The lower scales for the new physics in these
models could imply much larger effects for flavor changing processes
than were found in the previous analysis of the models without
T-parity. 

In the Littlest Higgs model with T-parity, the Standard Model
particles have even T-parity while the heavy gauge bosons and Higgs
triplet scalars are odd under T-parity.   To incorporate the T-parity
symmetry, the fermion sector becomes more complex \cite{6} with additional
fermions having both even and odd T-parity.   

Buras et al.\cite{11} consider two scenarios, one with a degenerate mirror
fermion spectrum corresponding to a minimal flavor violation (MFV)
scenario and one with a mirror fermion hierarchy different from CKM
corresponding to a non-MFV scenario.  A wide range of parameter space
is explored for the impact of the new physics contributions on
predictions for weak mixing, CP asymmetries and rare decays. 

Some general conclusions are possible.  There are now regions where 
the $B_s$ oscillation parameter, $\Delta M_s$, is smaller than the Standard Model value
in closer agreement with the recent Tevatron measurements \cite{12}.  The
$\sin2\beta$ ``problem'' can be solved, ie. the difference between the value $0.786
\pm 0.052$ from tree-level decays and the value $0.675\pm 0.026$ from the CP
asymmetries \cite{13}.  The two processes actually measure somewhat
different angles in the Littlest Higgs model.  The semileptonic CP
asymmetry, $A_{SL}^s$, can be enhanced by a factor of $10$--$20$ and $A^d_{SL}$ by a factor of
3 above the SM predictions.  The CP asymmetry, $S_{\psi\phi}$, can be
as high as $0.3$ 
compared to the Standard Model value of $0.04$.  The rare decays of
neutral and charged $K$-mesons to neutrinos can be enhanced by an order
of magnitude over the SM predictions.  In fact there are two branches
of parameter space, one where the branching ratio for $K^+ \to \pi^+
\nu\bar{\nu}$ can be greatly
enhanced but the decay $K_L\to \pi^0\nu\bar{\nu}$ is SM-like and one where $K_L\to\pi^0\nu\bar{\nu}$ can be greatly
enhanced with only modest enhancements of $K^+\to\pi^+\nu\bar{\nu}$.
I refer to the original 
papers \cite{10,11} for detailed plots. 

\section{Conclusions}
Little Higgs models offer a novel approach to resolving the tension
between the fine-tuning of the Higgs mass parameter and the
constraints of precision electroweak experiments.   The Littlest Higgs
model is a specific realization of these models which can address all
aspects of physics from the electroweak scale to scales of order $10$--$30$
TeV.  Flavor physics is highly constrained in the Standard Model by
the CKM structure of flavor-changing processes. 

Weak mixing and rare decay processes provide sensitive probes of the
new physics contained in the Littlest Higgs model.   Because of the
lower new physics scales possible in the Littlest Higgs model with
T-parity, large flavor-changing effects are possible in some regions
of parameter space providing specific opportunities for probing
physics beyond the Standard Model.  

A novel aspect of these calculations is the presence of log divergent contributions to the flavor-changing Z-boson vertex.   These contributions are gauge invariant and signal an additional sensitivity of flavor physics to the ultraviolet completion of the Littlest Higgs model.

\section*{Acknowledgements}
Talk presented at the 2006 International Workshop on The Origin of 
Mass and Strong Coupling Gauge Theories held November 21-24, 2006 
at Nagoya, Japan.  I would like to thank Koichi Yamawaki and the 
other members of the Organizing Committee for their hospitality
and for inviting me to participate in SCGT2006.

The transparencies for this talk were prepared by A. Poschenrieder
(Technical University, Munich) based on his lecture at the Ringberg
Workshop on Heavy Flavor Physics, October, 2006.  

Fermilab is operated by Fermi Research Alliance, LLC under Contract 
No. DE-AC02-07CH11359 with the United States Department of Energy.

\end{document}